\documentclass[sigconf]{acmart}
\AtBeginDocument{%
  }

\setcopyright{acmlicensed}
\copyrightyear{2018}
\acmYear{2018}
\acmDOI{XXXXXXX.XXXXXXX}
\acmConference[Conference acronym 'XX]{Make sure to enter the correct
  conference title from your rights confirmation email}{June 03--05,
  2018}{Woodstock, NY}
\acmISBN{978-1-4503-XXXX-X/2018/06}

\usepackage{bibentry}
\usepackage{multirow}
\usepackage{xcolor}
\usepackage{colortbl}
\usepackage{enumitem}
\usepackage{CJKutf8}
\usepackage{subfig}

\usepackage{graphicx}
\usepackage{stfloats}    
\usepackage{afterpage}    
\usepackage{lipsum}       
\usepackage{kantlipsum}   

\begin{document}

\title{OneLoc: Geo-Aware Generative Recommender Systems for Local Life Service}


\author{Zhipeng Wei}
\authornote{Equal contribution.}
\affiliation{%
\country{Kuaishou Inc., Beijing, China}
}
\email{weizhipeng@kuaishou.com}

\author{Kuo Cai}
\authornotemark[1]
\affiliation{%
\country{Kuaishou Inc., Beijing, China}
}
\email{caikuo@kuaishou.com}

\author{Junda She}
\affiliation{%
\country{Kuaishou Inc., Beijing, China}
}
\email{shejunda@kuaishou.com}

\author{Jie Chen}
\affiliation{%
\country{Kuaishou Inc., Beijing, China}
}
\email{chenjie20@kuaishou.com}

\author{Minghao Chen}
\affiliation{%
  \country{Kuaishou Inc., Beijing, China}
}
\email{chenminghao@kuaishou.com}

\author{Yang Zeng}
\affiliation{%
  \country{Kuaishou Inc., Beijing, China}
}
\email{zhengchengyi@kuaishou.com}

\author{Qiang Luo$^{\dag}$}
\affiliation{%
\country{Kuaishou Inc., Beijing, China}
}
\email{luoqiang@kuaishou.com}

\author{Wencong Zeng}
\affiliation{%
  \country{Kuaishou Inc., Beijing, China}
}
\email{zengwencong@kuaishou.com}

\author{Ruiming Tang$^{\dag}$}
\affiliation{%
\country{Kuaishou Inc., Beijing, China}
}
\email{tangruiming@kuaishou.com}

\author{Kun Gai}
\affiliation{%
\country{Unaffiliated, Beijing, China}
}
\email{gai.kun@qq.com}

\author{Guorui Zhou}
\authornote{Corresponding author.}
\affiliation{%
\country{Kuaishou Inc., Beijing, China}
}
\email{zhouguorui@kuaishou.com}

\renewcommand{\shortauthors}{Zhipeng Wei, Kuo Cai et al.}

\begin{abstract}
Local life service is a vital scenario in Kuaishou App, where video recommendation is intrinsically linked with store's location information. Thus, recommendation in our scenario is challenging because we should take into account user's interest and real-time location at the same time. In the face of such complex scenarios, end-to-end generative recommendation has emerged as a new paradigm, such as OneRec in the short video scenario, OneSug in the search scenario, and EGA in the advertising scenario. However, in local life service, an end-to-end generative recommendation model has not yet been developed as there are some key challenges to be solved. The first challenge is how to make full use of geographic information. The second challenge is how to balance multiple objectives, including user interests, the distance between user and stores, and some other business objectives. To address the challenges, we propose OneLoc. Specifically, we leverage geographic information from different perspectives: (1) geo-aware semantic ID incorporates both video and geographic information for tokenization, (2) geo-aware self-attention in the encoder leverages both video location similarity and user's real-time location, and (3) neighbor-aware prompt captures rich context information surrounding users for generation. To balance multiple objectives, we use reinforcement learning and propose two reward functions, i.e., geographic reward and GMV reward. With the above design, OneLoc achieves outstanding offline and online performance. In fact, OneLoc has been deployed in local life service of Kuaishou App. It serves 400 million active users daily, achieving $21.016\%$ and $17.891\%$ improvements in terms of gross merchandise value (GMV) and orders numbers.
\end{abstract}




\maketitle

\begin{figure}
    \centering
    \includegraphics[width=0.99\linewidth]{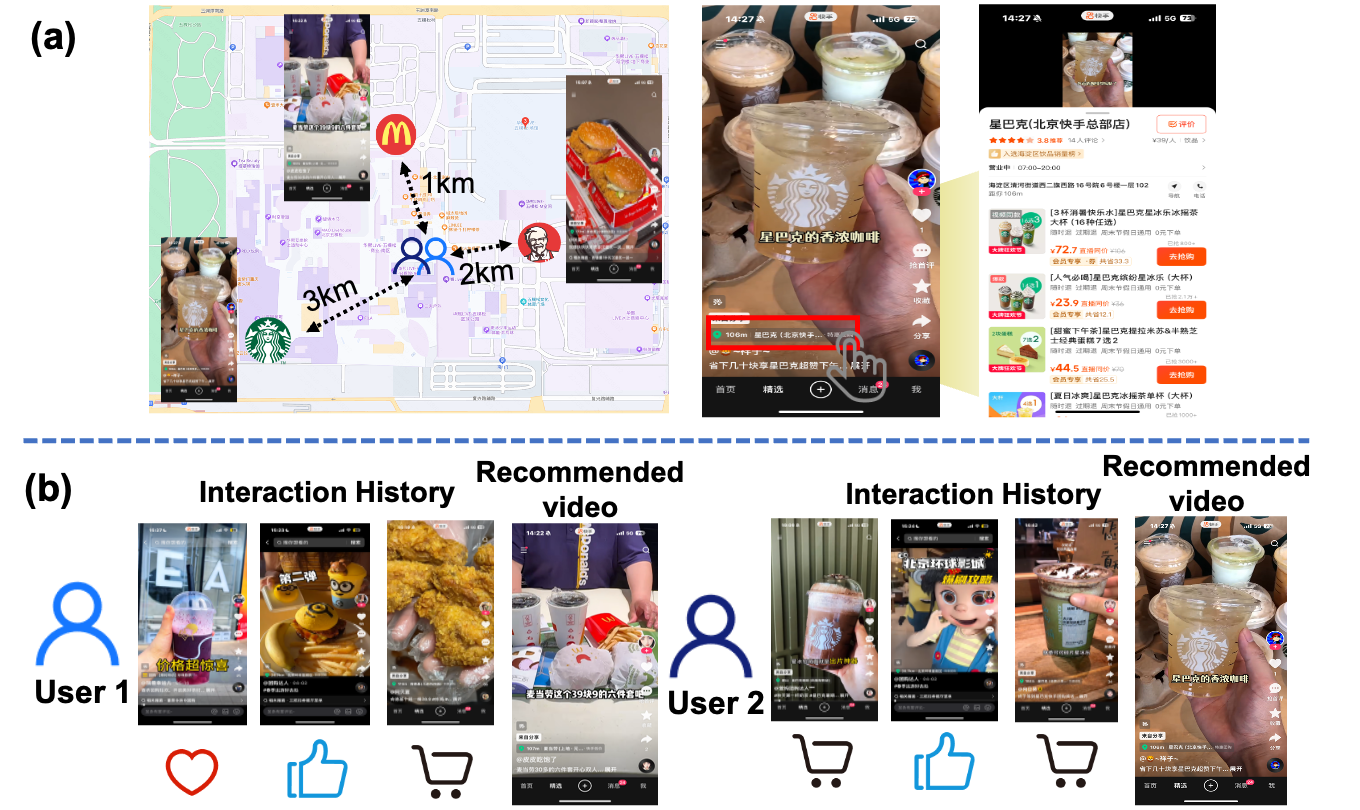}
    \caption{
    The background of local life service in Kuaishou. (a) In local life service, we recommend videos with stores’ location information displayed in the left-bottom (red box). After clicking, it will display detail store and products information. (b) Our system makes recommendation according to users' preference and the distance of the stores.
    }
    \label{fig:intro}
\end{figure}

\section{Introduction}

Local life service recommendation (LLSR)~\cite{lan2025benchmarkingLLS,wang2025fim} has become a critical scenario for some major internet platforms, such as Kuaishou, Meituan and Douyin.
In the local life service of Kuaishou App, videos with stores' location information displayed in the left-bottom corner are uploaded. 
LLSR aims to recommend such videos to near-by users to attract their consumption.  
For the ease of understanding, an illustrated example is presented in Figure~\ref{fig:intro}.
When a user enters the Kuaishou App, her real-time location is recorded\footnote{Local life service is disabled if the user forbids Kuaishou App to visit her real-time location.} and LLSR recommends videos with the consideration of both the user's interests and location information. 
For instance, at the same location, two users want to buy a drink and open Kuaishou App. 
Meanwhile in local life service scenario, there are three videos nearby, which are about McDonald's, KFC and Starbucks.
The first user recently took positive actions (such as sharing, liking, etc) on videos about KFC, and LLSR would recommend the video of McDonald's instead of KFC to the user, because McDonald's is closer.
The second user recently made multiple purchases of Starbucks and LLSR would recommend the video of Starbucks even though it is farther compared with KFC and McDonald's.
The above example indicates that user's interests and real-time location are both vital in our scenario. 

Recently, Large Language Models (LLMs)~\cite{openai2024gpt4technicalreport} based on autoregressive generation have demonstrated powerful capabilities in zero-shot learning and multi-domain generalization. This breakthrough success drives the emergence of a new paradigm of recommendation~\cite{guo2025onesug, zheng2025ega, deng2025onerec, zhou2025onerec-tech, liu2024end-to-end, zhai2025multimodal, zheng2024lc-rec, rajput2023tiger, zhai2024hstu, wang2024letter, hong2025eager, hou2025actionpiece, yang2025COBRA,wang2025poi_gr} - the shift from matching-based to generation-based approaches, known as generative recommendation. As an innovative paradigm, it demonstrates superior performance compared to traditional cascaded recommendation architectures across various industrial applications. In short video recommendation scenario, OneRec~\cite{zhou2025onerec-tech} proposes a generative recommendation paradigm with reward models to align with user preference and industry requirements.
In search scenario, Onesug~\cite{guo2025onesug} proposes an end-to-end generative framework for e-commerce query suggestion.
In advertising scenario, EGA~\cite{zheng2025ega} designs an end-to-end generative advertising system for critical advertising requirements, such as explicit bidding, creative selection, ad allocation, and payment computation. 
In POI recommendation scenario, GNPR-SID~\cite{wang2025poi_gr} mitigates the generative recommendation paradigm and achieves promising result. 
However, in the local life service scenario, generative recommendation models have not yet been developed as there are some key challenges that need to be solved.
The \emph{first} challenge is to make full use of geographic information. Existing works propose to utilize geographic from different aspects.
GNPR-SID~\cite{wang2025poi_gr} uses geographic information to construct semantic IDs, thus allocating each item a geography-aware representation.
Rotan~\cite{feng2024rotan} leverages temporal information as positional encoding, as part of input to attention mechanism.
TPG~\cite{luo2023tpg} proposes using temporal information as prompts to guide location recommendation.
As can be observed above, existing works utilize geographic information from either \emph{representation} perspective or \emph{decoder prompt} perspective. Such strategies are far from making full use of geographic information. 
The \emph{second} challenge is to balance multiple objectives, such as user interests, the distance between user and stores, and some other business objectives.
In the field of LLMs, ChatGPT~\cite{openai2024gpt4technicalreport}  uses reinforcement learning of human feedback (RLHF) to balance model's general capabilities and user experience.
In short video recommendation, Onerec~\cite{zhou2025onerec-tech} balances multiple recommendation objectives through carefully designed rewards.
However, how to balance multiple objectives in local life service has not yet been researched.

To address these two challenges, we propose OneLoc (\underline{One} Model for \underline{Loc}al Life Service), an end-to-end generative recommendation model tailored for the local life service scenario.
OneLoc is with encoder-decoder structure and trained in a two-stage paradigm, following~\cite{zhou2025onerec-tech}, i.e., pre-training and post-traninig (reinforcement learning).
To make full use of geographic information (w.r.t. the first challenge), we design three key components to enhance such information: (1) \emph{geo-aware semantic IDs} from \emph{representation} perspective, (2) \emph{geo-aware self-attention} from \emph{encoder attention} perspective and (3) \emph{neighbor-aware prompt} from \emph{decoder prompt} perspective.
\emph{Firstly}, traditional methods (such as Rotan~\cite{feng2024rotan} and TPG~\cite{luo2023tpg}), encode geographic information as an independent feature (represented as numeric IDs), which fails to model semantic relationships between similar locations.
Similar to GNPR-SID~\cite{wang2025poi_gr}, we represent videos by Semantic IDs (SIDs) which are tokens derived from textual video descriptions and other features. Geographic information is integrated into the raw video features such that SIDs of videos contain geographic semantics, for which we refer to \emph{geo-aware SIDs}.
\emph{Secondly}, we propose \emph{geo-aware self-attention} structure to extract user behavior sequential patterns with geographic information encoded. To the best of our knowledge~\cite{zhang2025poi-survey}, this is the first architecture to model these two kinds of information simultaneously.
Geo-aware self-attention calculates attention scores with the user's interacted video sequence and the corresponding location sequence.
More specifically, the attention score between two videos is defined as the combination of two parts: (1) content similarity using comprehensive video embedding (with video information and its geographic information) as Queries and Keys and (2) location context similarity between the two videos to further enhance the location semantics.
\emph{Thirdly}, existing methods~\cite{luo2023tpg,feng2024rotan,wang2025poi_gr} utilize user's coordinates or timestamps as prompts to guide generating recommendation.
Besides user's own location, we additionally utilize surrounding location information to model richer user's geographic context.
More specifically, \emph{neighbor-aware prompt} is constructed in decoder to guide generation by modeling user's real-time location as Query and surrounding location as Key and Value such that neighborhood information can be effectively integrated into the user's location.

To balance multiple objectives (w.r.t. the second challenge), we design a reward function, including both geographic reward signal and GMV reward signal for reinforcement learning.
In LLSR scenario, the distance between a user and stores in the videos is a critical factor in determining the user's consumption in these stores. Therefore, we design a geographic reward function, where a closer location would get a higher reward.
Besides distance, GMV is another vital objective, therefore we further design the GMV reward function. Finally, the geographic reward function and GMV reward function are used to guide reinforcement learning.

To summarize, our contributions are as follows:
\begin{itemize}[leftmargin=*]
\item We propose OneLoc, an end-to-end generative recommender system for short-video local life service. The framework integrates a generative architecture with a business-value-optimized reinforcement learning module, significantly outperforming traditional cascading recommendation models.
\item We propose three core components to make full use of geographic information: (1) a geo-aware tokenizer (generating geo-aware SIDs) that combines geographic semantics with multi-modal video information, (2) a geo-aware self-attention structure which captures user behavior sequential patterns with geographic information encoded and (3) a neighbor-aware prompt in decoder to guide recommendation generation which considers user's location as well as neighborhood information. 
\item In reinforcement learning phase, we propose two reward functions:
(1) Geographic Reward to reinforce the distance factor between a user and stores in videos and (2) GMV Reward for business objectives.
\item Extensive offline experiments on Kuaishou large-scale industry dataset demonstrate the effectiveness of OneLoc, compared to traditional models (including generative models). Ablation studies are also conducted which validates the functionalities of our proposed components. 
\item OneLoc has been deployed in local life service of Kuaishou App and achieves $21.016\%$ and $17.891\%$ improvements in terms of GMV and order numbers. Now OneLoc serves the full traffic in our system, supporting $400$ million users in local life service daily.
\end{itemize}

\begin{figure*}[htbp]
    \centering
    \includegraphics[width=0.98\linewidth]{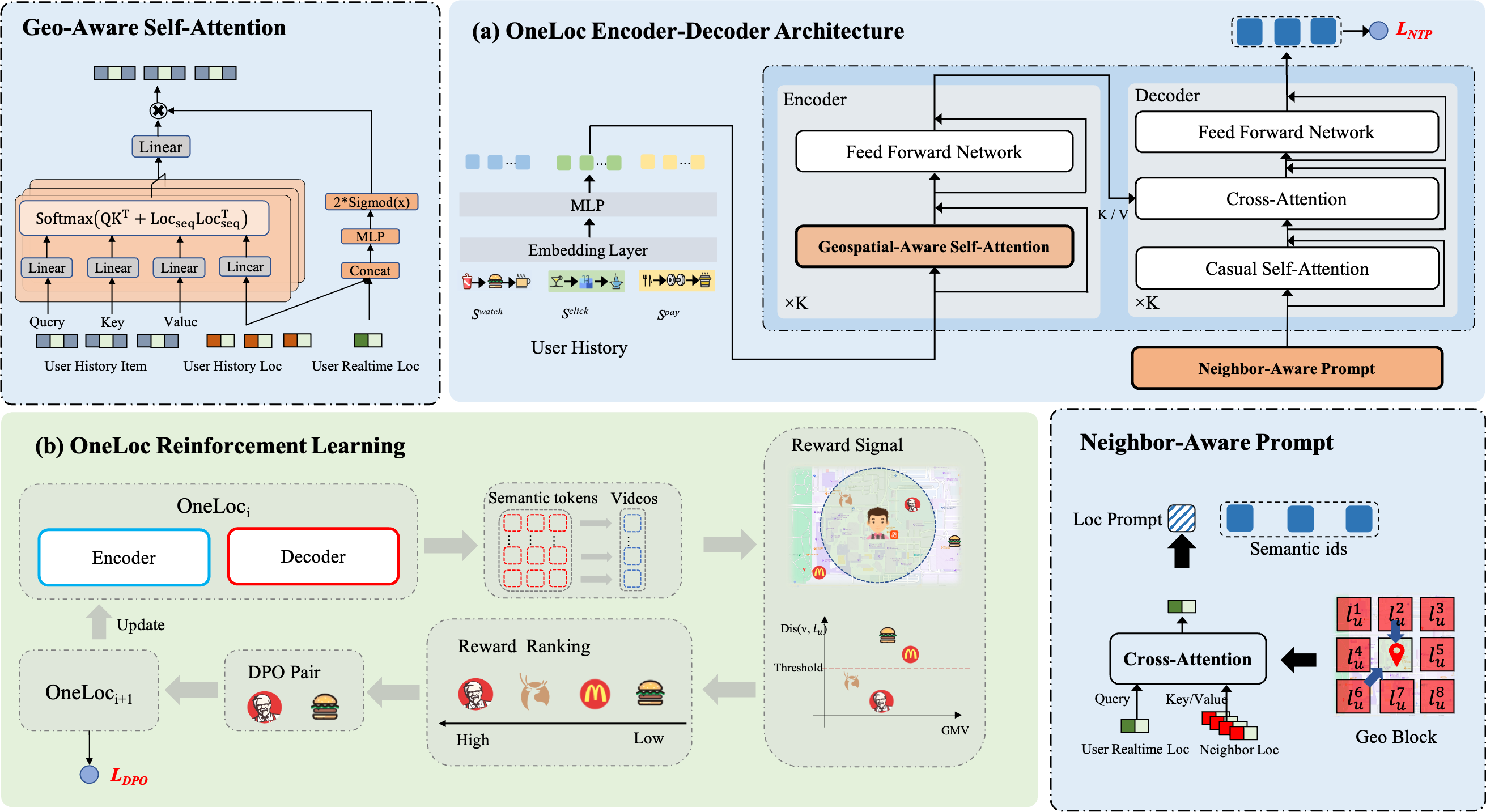}
    \caption{The overall framework of OneLoc includes encoder-decoder architecture and reinforcement learning. (a) The core module in encoder is geo-aware self-attention, which introduces location similarity into attention scores and leverages user's real-time location as a gate to control the outputs. The outputs of the encoder act as keys and values of the cross-attention in the decoder. Meanwhile, a neighbor-aware prompt is use to control the generation, which aggregates the surrounding context using cross-attention. Finally, the output of the decoder is used to calculate the next token prediction loss. (b) With the model parameters initialized , we sample multiple videos from the probability distribution. After that, we scores the videos using the geographic reward and GMV reward. Finally, we choose videos with the highest score and lowest score as preference pair for direct preference optimization.} 
    \label{fig:model}
\end{figure*}

\section{Method}

In this section, we first elaborate on the problem formulation in our scenario.
After that, we introduce how we leverage geographic information in an encoder-decoder architecture, including core designs geo-aware tokenizer, geo-aware self-attention in the encoder and neighbor-aware prompt in the decoder.
In addition, we elaborate on how we design the geographic reward and the GMV reward in reinforcement learning.
In the end, we introduce the final loss of our framework.
The overall framework is depicted in Figure~\ref{fig:model}.

\subsection{Problem Formulation}
Let $\mathcal{U}$, $\mathcal{V}$ and $\mathcal{L}$ represent the set of users, videos, and location, respectively.
In our task, each video $v \in \mathcal{V}$ is assigned a location $l \in \mathcal{L}$.
The location is actually a GeoHash block and contains rich context information, including geographical coordinates, brand, and category.
Thus, each video can be represented by a video embedding $e^{v}$, a location id embedding $e^{lid}$ and a location context embedding $e^{lc}$.
In our scenario, the location $l_u$ where a user interacts with videos is also associated with a context embedding $e^{lc}_u$. 
Furthermore, each video is mapped to a semantic ID, denoted by $Q_v=(q^1_v,q^2_v,...,q_v^{T})$, where $T$ is the number of codebooks.
Given a user's interacted video sequence $S=\{v_1,v_2,...,v_{t-1}\}$ and real-time location $l_u$, the objective is to predict the next video $v_t$ that would attract consumption, which is formulated as maximize $p(v_{t}|l_u,S)$.

\subsection{Geo-aware Semantic IDs}
Following OneRec~\cite{zhou2025onerec-tech}, we use res-kmeans to generate semantic IDs, which employs residual quantization to map video embedding into multi-level discrete codes.
To inject geographic information, each residual $r^0_i$ in the initial residual set $\mathcal{R}^0$ is represented as an embedding of video content and location context information, extracted by a multimodal large language model.
At each layer $i$, we construct the codebook $\mathcal{C}^i$ by applying K-means clustering on the residuals $\mathcal{R}^i$:
\begin{align}
    \mathcal{C}^i &= \textit{K-means}(\mathcal{R}^i,N_c), 
\end{align}
where the codebook $\mathcal{C}^i=\{c^i_k|k=1,2,...,N_c\}$ is the centroids set derived from \textit{K-means}, $N_c$ is the codebook size.

With the codebook $\mathcal{C}^i$, we can compute the nearest centroid index $q^{i}_j$ for each residual $r^i_j$ in the residual set $\mathcal{R}^i$.
The residual $r^{i+1}_j$ of the next layer is the difference between $\mathcal{R}^i_j$ and its nearest centroid:
\begin{align}
    & q^{i}_j = \underset {k} { \operatorname {arg\,min} } ||c^{i}_k-r^{i}_j|| , \\
    & r^{i+1}_j = r^i_j-c^{i}_{q^{i}_j},
\end{align}
where the index $q^{i}_j$ is the $i$-th code of the semantic ID.

The above process iterates $T$ times and we can get $T$ codebooks $(\mathcal{C}^1,\mathcal{C}^2,...,\mathcal{C}^T)$.
In our scenario, we set $T$ to $3$.
Thus, we can transform the target video $v_{t}$ into its corresponding semantic ID $Q_t=\textit{Tokenize}(v_{t})=(q^1_t,q^2_t,q^{3}_t)$.

\subsection{Encoder}

The encoder is used to encode the user behavior sequence for useful information extraction.
In this section, we first introduce the input and architecture of the encoder.
After that, we elaborate on the geo-aware self-attention, which is the core module to capture user's behavior patterns.

\subsubsection{Multi-behavior Sequence.} 
In our scenario, watching sequences are crucial for capturing user preference.
To capture different scales of user behavior patterns, we additionally incorporate user clicking and purchasing sequence:
\begin{align}
    S = \{S^{watch}, S^{click}, S^{pay}\},
\end{align}
where $S^{watch}$, $S^{click}$ and $S^{pay}$ are the watching, clicking and purchasing sequence, respectively.

The multi-behavior sequence $S$ is then transformed into a comprehensive embedding sequence $Z$ and a location context embedding sequence $E^{lc}$, which are sent to the encoder:
\begin{align}
    & Z = \{z_1, z_2,...,z_i,..., z_{|S|}\}, \\
    \label{Z}
    & z_i = MLP(Concat(e^{v}_i,e^{lid}_i,e^{lc}_i)), \\
    & E^{lc} = \{e^{lc}_1, e^{lc}_2,...,e_i^{lc},...,e^{lc}_{|S|}\},
\end{align}
where $|S|$ is the length of sequence $S$, \textit{Concat} is the concatenate operation along the feature dimension, $z_i$ is the video embedding, $e^{v}_i$, $e^{lid}_i$ and $e^{lc}_i$ are the video id embedding, location id embedding and location context embedding of the $i$-th video in the sequence, respectively, $E^{lc}$ is the location context embedding sequence.

\subsubsection{Encoder architecture.}
The embedding sequences mentioned above, together with the user location, are then sent to the encoder, which stacks $K$ transformer blocks. 
Each block contains a $\textit{GA-Attn}$ (Geo-aware Self-attention) module and a $\textit{FFN}$ module with $\textit{RMSNorm}$.
Formally:
\begin{align}
    & \widetilde{Z}^{i+1} = Z^i + \textit{GA-Attn}(RMSNorm(Z^i),E^{lc}, e^{lc}_u), \\
    & Z^{i+1} = \widetilde{Z}^{i+1} + FFN(RMSNorm(\widetilde{Z}^{i+1})), \\
    & Z^0 = Z,
\end{align}
where $Z$ is the input embedding sequence of user $u$, $Z^{i}$ is the output of the $i$-th encoder layer, $e^{lc}_u$ is the context embedding of user's location $l_u$, \textit{GA-Attn} is the core module to capture user behavior patterns.

\subsubsection{Geo-aware Self-attention.}
\textit{GA-Attn} module is proposed to capture relevant behaviors from the user's interaction history according to the user's real-time location.
Specifically, the attention score is defined as the combination of two parts: (1) comprehensive similarity using the comprehensive embedding sequence $Z$ as Queries and Keys, and (2) location context similarity using location context embedding sequence $E^{lc}$ to further enhance location semantics.
\begin{align}
    A &=\textit{Softmax}((ZW_q)(ZW_k)^T/\sqrt{d}+ E^{lc}(E^{lc})^T), \label{eq:location context scores} \\
    \widetilde{O} &= A(ZW_v)W_o,
    \label{eq:gate}
\end{align}
where $W_q, W_k, W_v, W_o$ are the weight matrix of query, key, value and output, $E^{lc}$ is the location context sequence. 

To inject user's real-time location information, we further leverage the location context embedding $e_u^{lc}$ as a gate: 
\begin{align}
    &g_i = 2*\textit{Sigmoid}(MLP(Concat(e^{lc}_u,E^{lc}_{i*}))), \\
    \label{eq:gate}
    &O_{i*} = g_i \widetilde{O}_{i*},
\end{align}
where $E^{lc}_{i*}$ is the $i$-th row of $E^{lc}$, $g_i\in (0,2)$ is the scaling parameter, $O_{i*}$ is the $i$-th row of the output of \textit{GA-Attn} module.

\subsection{Decoder}

The decoder is used to generate recommendation results according to the output of the encoder and the prompt about the user's location.
To model user's richer geographic context, we propose neighbor-aware attention, which additionally utilizes user's surrounding location information.
The neighbor-aware attention together with the semantic ID is then sent to the decoder to compute the next token prediction loss.


\subsubsection{Neighbor-aware Prompt}
\label{geo-prompt}
In local life service, it is vital to model location contexts surrounding the user, such as surrounding brands, bestselling products, etc.
Thus, we use cross-attention to capture the surrounding information.
Specifically, given the user's real-time geographic location $l_u$, we calculate the surrounding geographic location $\{l^1_u,l^2_u,...,l^8_u\}$ and obtain their context information.
Thus, we can obtain their context embedding $e^{lc}_{u}$ and $E^s=\{e^{lc}_{l^1_u},...,e^{lc}_{l^8_u}\}$.
After that, we calculate cross attention using $e^{lc}_{u}$ as query and $E^s$ as keys and values:
\begin{align}
    e^s &= CrossAttn(e^{lc}_{u}, E^s),
\end{align}
where $E^s$ is the context embedding set of surrounding locations, $e^{lc}_{u}$ is the context embedding of the user's real-time location, $e^s$ is the embedding of neighbor-aware prompt, $CrossAttn(e^{lc}_{u}, E^s)$ calculates cross-attention using $e^{lc}_{u}$ as query and $E^s$ as key and value.

\subsubsection{Decoder architecture.}
The decoder stacks $K$ blocks, each of which contains a casual self-attention module, a cross-attention module and a feed forward network (\textit{FFN}) module with $\textit{RMSNorm}$.
Formally, to generate the semantic ID of the target video, the calculation can be represented as follows:
\begin{align}
    & \widetilde{H}^{i+1} = H^i + SelfAttn(RMSNorm(H^i)), \\
    & \widetilde{H}^{i+1} = \widetilde{H}^{i+1} + CrossAttn(RMSNorm(\widetilde{H}^{i+1}),S^K), \\
    & H^{i+1} = \widetilde{H}^{i+1} + FFN(RMSNorm(\widetilde{H}^{i+1})), \\
    & H^0 = \{e^{s}, e_{q^1_t},e_{q^2_t},e_{q^3_t}\},
\end{align}
where $H^{i}$ is the output of the $i$-th decoder layer, $\widetilde{H}^{i+1}$ denote the intermediate results of the $i+1$-th decoder layer, $CrossAttn(a,b)$ calculates cross-attention using $a$ as query and $b$ as key and value, $SelfAttn$ and $FFN$ is the self-attention module and feed forward network module, $e^{s}$ is the embedding of geo-aware prompt, $e_{q^1_t}$, $e_{q^2_t}$ and $e_{q^3_t}$ are the embedding of three-digit semantic ID of the target video.

The output $H^{K}$ of the decoder is used to predict the next token.
Specifically, the output embedding $H^{K}_0$ of geo-aware prompt is used to predict the first digit of the target semantic ID $q_t^1$.
The output embedding $H^{K}_1$ of $q_t^1$ is used to predict the second digit of the target semantic ID $q_t^2$.
Formally, the training loss is the cross-entropy loss:
\begin{align}
    &\hat{y}^j = \textit{Softmax}(\textit{MLP}(H^K_{j-1}))\in \mathbb{R}^{N_c}, j\in\{1,2,3\}\\
    &L_{ntp} =  \sum_{j=1}^3 -log\hat{y}^j_{q_t^j},
\end{align}
where $\hat{y}^i_t \in \mathbb{R}^{N_c}$ is the predicted probability distribution of the $i$-th digit of the target semantic ID, $\hat{y}^i_{q_t^i}$ is the predicted probability of $q_t^i$.

After training a certain number of samples, we obtain a pre-trained model with parameters $\theta_0$.
Using the interacted sequence $S$ as input of the encoder and the real-time location $l_u$ as input of the decoder, the model outputs the probability of semantic ID. 
Formally, $p_{\theta_0}(Q_t|S,l_u)=p_{\theta_0}(q_t^1|S,l_u)p_{\theta_0}(q_t^2|S,l_u,q_t^1)p_{\theta_0}(q_t^3|S,l_u,q_t^1,q_t^2)$.

\subsection{Reinforcement Learning}

In the pre-training phase above, the model only fits the exposed videos through next token prediction, which are obtained from the traditional recommendation system.
Although the exposed videos satisfy multiple objectives to some degree, it is hard to balance multiple objectives in a fine-grained manner if only aligning with exposed videos.
To solve this challenge, we introduce two rewards tailored for our scenario, i.e., geographic reward and GMV reward.
After that, we use direct preference optimization (DPO) for alignment.

\subsubsection{Reward Signals.}

The geographic reward signal is proposed to encourage generating videos nearby and thus a closer video would get a higher reward:
\begin{align}  
    R^{geo}(v,l_u) = \begin{cases}
    0 & \text{if } Dis(v,l_u)>D, \\
    \frac{1}{Dis(v,l_u)} & \text{else } .
    \end{cases}
\end{align}
where $\textit{Dis}(v,l_u)$ operation would calculate the distance between video $v$ and the user's location $l_u$, $D$ is the distance threshold.

The GMV reward signal is proposed to encourage the generation of videos that attract consumption.
To achieve this goal, we use a traditional GMV scoring model as the reward model:
\begin{align}
    R^{gmv}(v,S,l_u) = GMV(v,S,l_u),
\end{align}
where \textit{GMV} is a GMV scoring model.

\begin{table*}[htbp]
\centering
\caption{Recommendation performance improvement on different datasets in terms of Recall and NDCG. The best and second-best results are highlighted in \textbf{bold} font and \underline{underlined}. The superscript * indicates the improvement is statistically significant where the p-value is less than 0.05.
}
\resizebox{1.0\linewidth}{!}{
\begin{tabular}{clccccccccccc}
\hline

\multirow{2}{*}{Dataset} & \multicolumn{1}{l}{\multirow{2}{*}{Metric}} & \multicolumn{5}{c}{Traditional} & \multicolumn{2}{c}{POI Related} & \multicolumn{3}{c}{Generative} & \multirow{2}{*}{Improvement} \\ \cmidrule(lr){3-7} \cmidrule(lr){8-9} \cmidrule(lr){10-12}
\multicolumn{2}{c}{} & SASRec & BERT4Rec & GRU4Rec & Caser & $\text{S}^3$-Rec & TPG & Rotan & TIGER    & GNPR-SID   & Ours   &  \\ \hline
\multirow{6}{*}{KuaiLLSR}  &Recall@5&0.0927&0.0682&0.0350&0.0438&0.1218&0.1750&0.2185&0.2832&\underline{0.3142}&\textbf{0.3565}*&13.46\%\\ 
&Recall@10&0.1336&0.1071&0.0602&0.0712&0.1889&0.2559&0.2843&0.3637&\underline{0.4207}&\textbf{0.4563}*&8.46\%\\ 
&Recall@20&0.2048&0.1623&0.1090&0.1147&0.2808&0.3241&0.3563&0.4413&\underline{0.5056}&\textbf{0.5584}*&10.44\%\\ 
&NDCG@5&0.0408&0.0311&0.0178&0.0221&0.0793&0.0897&0.1029&0.1500&\underline{0.1775}&\textbf{0.2032}*&14.47\%\\ 
&NDCG@10&0.0523&0.0338&0.0255&0.0266&0.0971&0.1018&0.1147&0.1584&\underline{0.1874}&\textbf{0.2114}*&12.81\%\\ 
&NDCG@20&0.0705&0.0565&0.0360&0.0369&0.1143&0.1117&0.1266&0.1615&\underline{0.1904}&\textbf{0.2151}*&12.97\%\\ 
\hline

\multirow{6}{*}{NYC}  &Recall@5&0.3151&0.2857&0.1977&0.2883&0.3071&0.3551&0.4448&0.4965&\underline{0.5311}&\textbf{0.6107}*&14.98\%\\ 
&Recall@10&0.3896&0.3564&0.2460&0.3570&0.3854&0.4441&0.5223&0.5514&\underline{0.5942}&\textbf{0.6563}*&10.45\%\\ 
&Recall@20&0.4506&0.4130&0.2889&0.4135&0.4503&0.5121&0.5834&0.6001&\underline{0.6455}&\textbf{0.6977}*&8.09\%\\  
&NDCG@5&0.2224&0.2074&0.1442&0.2044&0.2235&0.2464&0.3471&0.4131&\underline{0.4430}&\textbf{0.5355}*&20.88\%\\ 
&NDCG@10&0.2467&0.2304&0.1599&0.2267&0.2489&0.2755&0.3723&0.4276&\underline{0.4634}&\textbf{0.5504}*&18.77\%\\ 
&NDCG@20&0.2622&0.2448&0.1708&0.2410&0.2654&0.2927&0.3878&0.4443&\underline{0.4766}&\textbf{0.5608}*&17.66\%\\ \hline

\multirow{6}{*}{TKY}  &Recall@5&0.3450&0.2649&0.2514&0.3257&0.3365&0.3725&0.4333&0.5031&\underline{0.5354}&\textbf{0.5964}*&11.39\%\\ 
&Recall@10&0.4284&0.3326&0.3106&0.4067&0.4115&0.4601&0.5113&0.5808&\underline{0.6130}&\textbf{0.6620}*&7.99\%\\ 
&Recall@20&0.4976&0.3943&0.3651&0.4758&0.4739&0.5291&0.5894&0.6431&\underline{0.6675}&\textbf{0.7152}*&7.15\%\\  
&NDCG@5&0.2384&0.1907&0.1833&0.2273&0.2423&0.2591&0.3293&0.4003&\underline{0.4437}&\textbf{0.4961}*&11.81\%\\ 
&NDCG@10&0.2655&0.2127&0.2025&0.2535&0.2666&0.2881&0.3568&0.4251&\underline{0.4623}&\textbf{0.5174}*&11.92\%\\ 
&NDCG@20&0.2831&0.2284&0.2163&0.2711&0.2825&0.3051&0.3739&0.4401&\underline{0.4788}&\textbf{0.5306}*&10.82\%\\ \hline

\end{tabular}

}
\label{table:main}
\end{table*}

\subsubsection{Direct Preference Optimization.}
To construct preference pairs for direct preference optimization, we first sample several videos from the distribution of the pre-trained model $p_{\theta_0}$.
Specifically, we generate $N$ different videos for each sample $(S,l_u)$ by beam search:
\begin{align}
    \mathcal{B}_u^N = TopN(p_{\theta_0}(S,l_u)),
\end{align}
where \textit{TopN} operation would select $N$ videos with the highest probability, $\mathcal{B}_u^N$ is the generated results.

Then we calculate the reward for each of the results through $R^{geo}(v,l_u)$ and $R^{gmv}(v,S,l_u)$.
After that, we construct the preference pairs $D_{pairs}=(v_p,v_n,S,l_u)$ by choosing the video $v_p$ with highest reward as the positive sample and the video $v_n$ with lowest reward as the negative sample.
Given the preference pairs, we can now train a new model with parameters $\theta_{i+1}$, which is initialized from $\theta_{i}$.
The loss corresponding to each preference pair is as follows:
\begin{align}
    L_{dpo}= -log (\beta log\frac{p_{\theta_{i+1}}(Q_p|S,l_u)}{p_{\theta_{i}}(Q_p|S,l_u)} -\beta log \frac{p_{\theta_{i+1}}(Q_n|S,l_u)}{p_{\theta_{i}}(Q_n|S,l_u)}),
\end{align}
where $Q_p=Tokenize(v_p), Q_n=Tokenize(v_n)$ is the semantic ID of the positive video and the negative video, respectively.

The training loss in reinforcement learning is:
\begin{align}
    L = L_{ntp}+\lambda L_{dpo}.
\end{align}

\begin{figure}[ht]
    \centering
    \includegraphics[width=0.99\linewidth]{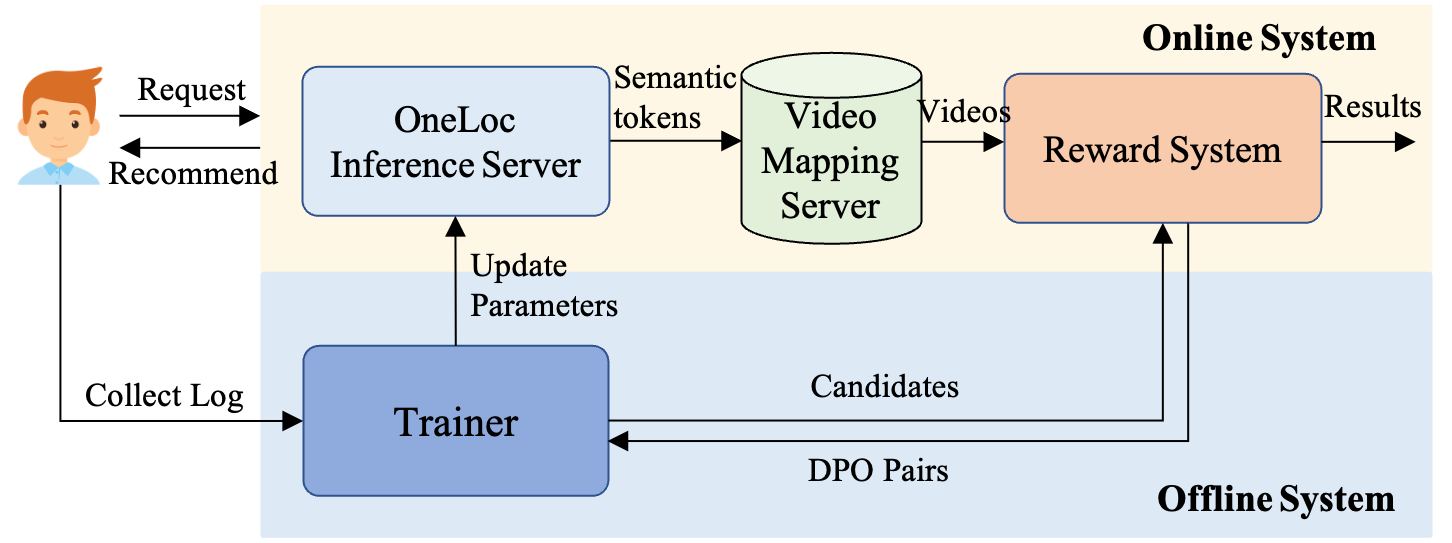}
    \caption{Framework of System Deployment}
    \label{fig:system}
\end{figure}

\section{System Deployment}
OneLoc has been successfully implemented in real-world local life service scenario of Kuaishou.
As illustrated in Fig.\ref{fig:system}, our deployment architecture consists of several core components: (1) Trainer, (2) OneLoc Inference Server, (3) Video Mapping Server, (4) Reward System. In the offline training phase, the Trainer collects user logs for streaming training, requests the Reward System to score and construct positive/negative sample pairs for RL training, and periodically updates parameters to the inference server during the process. The OneLoc inference server processes user requests by converting user features and real-time geographic locations into user tokens and geo prompts, which are then fed into the model for semantic token generation via beam search. The Video Mapping Server serves as a storage service that maps semantic tokens to video IDs. The candidate videos are subsequently processed by the Reward System through GMV score estimation and rule-based filtering, with the TopK results ultimately being recommended to users. For inference performance optimization, we implement mixed-precision computation, KV cache, dynamic batching, and TensorRT acceleration on NVIDIA A10 GPUs, achieving 25$\%$ Model FLOPs Utilization (MFU).

\section{Experiments}

In this section, we conduct extensive experiments on Kuaishou online platform to answer the following research questions:
\begin{itemize}[leftmargin=*]
\item \textbf{RQ1}: How does our proposed OneLoc perform compared to the state-of-the-art methods?
\item \textbf{RQ2}: How do the components of OneLoc (e.g., neighbor-aware prompt and geo-aware self-attention) affect the performance?
\item \textbf{RQ3}: How does OneLoc perform under different hyperparameters (i.e., model size, sequence length, loss weight $\lambda$)?
\end{itemize}

\subsection{Experimental Setting}

\begin{table}[ht]
\centering
\caption{Dataset statistics.}
\begin{tabular}{lccccc}
\hline
 & KuaiLLSR & NYC & TKY \\
\hline
\#users      & 60 M   & 1,042   & 2,187 \\
\#items  & 900 K  & 36,359  & 59,472 \\
\#interactions  & 440 M & 212,347 & 545,301 \\
\hline
\end{tabular}
\end{table}

\subsubsection{Datasets.}
The Kuaishou platform facilitates daily engagement of hundreds of millions of users with short videos about local life service, generating over hundreds of millions interactions. We construct the \textbf{KuaiLLSR} Dataset from this behavioral data for model training and evaluation. Specifically, the KuaiLLSR dataset contains eight days of uniformly sampled local life service short video interaction records, comprising 60 million users, 900K items, and 440 million interactions, with an average user sequence length of around 200. We train OneLoc in a streaming setup where the first 7 days' data are used for model training and the final day's data is reserved for evaluation. In addition, we conduct experiments on the publicly available \textbf{Foursquare} dataset, which contains user check-in records across various points of interest (POIs) in multiple cities. Each record includes the user ID, POI ID, geographical coordinates, and timestamp. We adopt LibCity’s~\cite{wang2021nyc_and_tky} standardized pre-processing pipeline to filter users and POIs, and then sort each user’s check-ins chronologically to construct interaction sequences. Following common practice, we adopt the NYC and Tokyo subsets, which differ in user density and POI distribution, allowing us to evaluate model performance under diverse spatial–temporal patterns. For these public datasets, we split the interaction sequences into 80\% for training, 10\% for validation, and 10\% for testing. Table 2 gives a rough sketch of the statistics of the five datasets.

\subsubsection{Baselines.}
We compare OneLoc with competitive baselines within two groups of work, traditional recommender models and generative recommender models: 1) \textbf{SASRec}~\cite{kang2018sasrec} employs a unidirectional Transformer encoder that effectively models user preference through self-attention mechanisms. 2) \textbf{BERT4Rec}~\cite{sun2019bert4rec} leverages BERT's pre-trained language representations to encoder user behavior sequences. 3) \textbf{GRU4Rec}~\cite{hidasi2015session} employs gated recurrent units to capture temporal dependencies within user interaction sequences for session-based recommendation tasks. 4) \textbf{S\textsuperscript{3}-Rec}~\cite{zhou2020s3} mines self-supervised learning with mutual information maximization for extracting inherent correlations within user behavior sequences. 5) \textbf{TPG}~\cite{luo2023tpg} is a transformer-based approach that leverages target timestamps as prompts to enhance geography-aware location recommendations. 6) \textbf{Rotan}~\cite{feng2024rotan} encodes time intervals through rotational position vector representations in transformer architectures, effectively capturing temporal dynamics in user behavior sequences. 7) \textbf{TIGER}~\cite{rajput2023tiger} pioneers a codebook-based semantic quantization framework through RQ-VAE, generating discrete code sequences as identifiers. 8) \textbf{GNPR-SID}~\cite{wang2025poi_gr} transforms POI information into discrete semantic identifiers and employs a generative approach for next-POI prediction.

\subsubsection{Evaluation Metrics.}
Following most prior works, we use two metrics for offline experiments: top-K Recall (Recall@K) and NDCG (NDCG@K). For the online A/B experiments, we use GMV and order quantity as core evaluation metrics.

\subsubsection{Implement Details.}
We train OneLoc using the AdamW optimizer (initial learning rate: $2\times10^{-4}$, weight decay: 0.1) on NVIDIA A800 GPUs. Each codebook layer uses K = 8192 clusters for semantic identifier clustering, with a total of L = 3 codebook layers. Both the encoder and the decoder stack 4 blocks with 1024 hidden units, 8 attention heads, and the dimension of the feedforward network (FFN) is 4096. The lengths of watch sequence, click sequence, and pay sequence are 256, 32, and 10 respectively. The DPO loss weight is set to $0.05$.

\subsection{Overall Performance (RQ1)}
To demonstrate the effectiveness of our method, we conducted experiments on two industry datasets (Foursquare and Kuaishou).
We compare our method with the state-of-the-art recommendation methods and the results are demonstrated in Table.\ref{table:main}.
From the result, we have the following observation: 
\begin{itemize}[leftmargin=*]
    \item (1) We observe that OneLoc consistently surpasses existing baselines in performance. Notably, on the KuaiLLSR datasets, OneLoc achieves multi-scale improvements—including a 13.46\% increase in Recall@5, a 10.44\% increase in Recall@20, a 14.47\% increase in NDCG@5, and a 12.97\% increase in NDCG@20, while also demonstrating superiority over the second-best baseline. Similarly, on the Foursquare global dataset, it attains an average performance boost of 13.18\% in Recall@5 and 16.34\% in NDCG@5. These improvements underscore the efficacy of our proposed integrates geographic generative architecture in the Local Life Service recommendation task. This is attributed to our novel approach, which leverages a dedicated geographic information module. The module integrates users' historical behavioral preferences with real-time spatial context to generate optimized recommendations.
    \item (2) All generative methods (Ours, TIGER, GNPR-SID) consistently exceed traditional recommender models by more than 29\% in Recall@5 and more than 45\% in NDCG@5. This improvement demonstrates the comprehensive semantic expression and deep reasoning capabilities of large models. Specifically, it significantly outperforms traditional recall solutions based on representation learning and ANN retrieval.
\end{itemize}

\subsection{Ablation Study (RQ2)}
In this section, we conduct an ablation study to investigate whether each component of our framework is effective (Recall and NDCG).
We focus on three key modules: (1) neighbor-aware prompt, (2) geo-aware self-attention, and (3) rewards function. 

\begin{figure}[ht]
    \centering
    \includegraphics[width=0.95\linewidth]{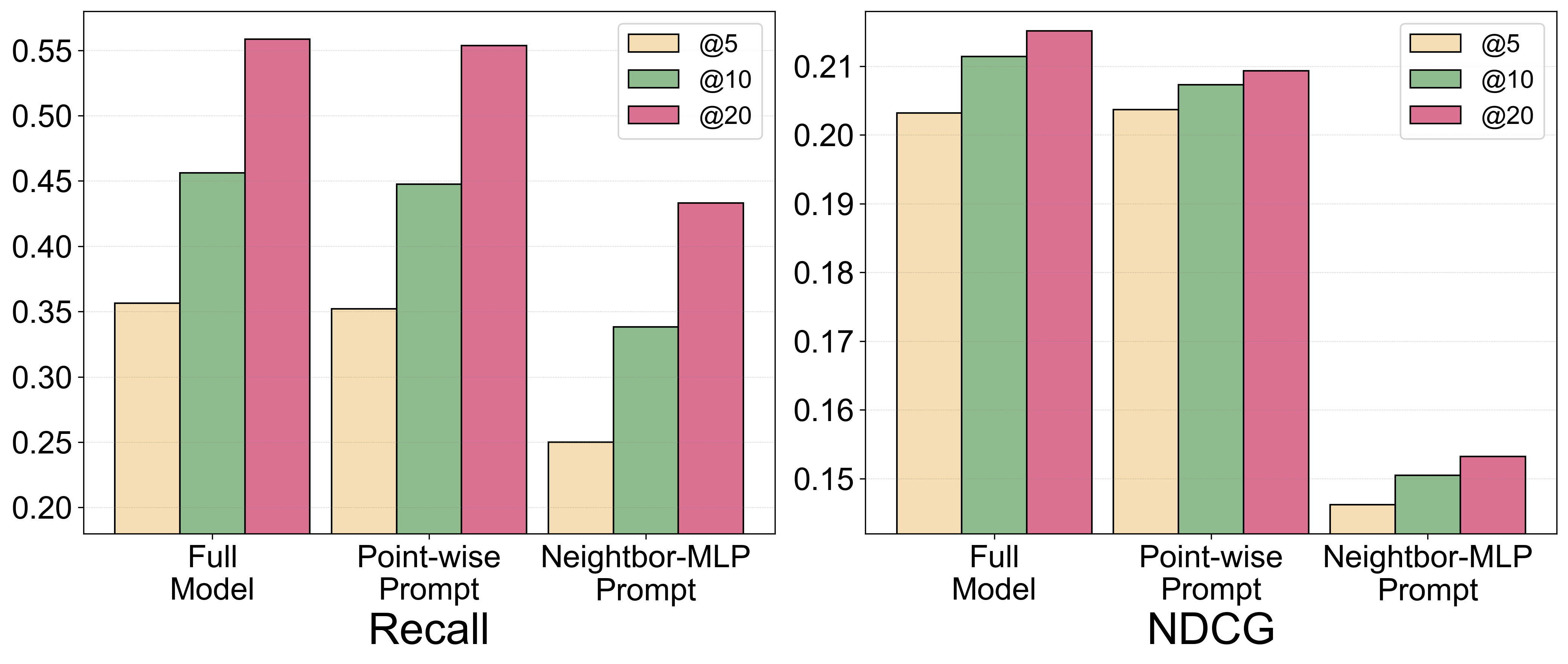}
    \caption{Ablation study of different prompt techniques. The result shows that Neighbor-aware prompt perform significantly better compared to Point-wise prompt and Neighbor-mlp prompt.}
    \label{fig:bar_ablation_study}
\end{figure}

\textbf{Neighbor-aware prompt.}
Neighbor-aware prompt uses cross attention to capture location context surrounding users.
Two questions may arise: (1) whether the surrounding context is effective and (2) whether cross attention is necessary to aggregate surrounding context.
To address the first question, we replace the neighbor-aware prompt with a point-wise prompt, which only uses the current location context rather than the surrounding context.

To address the second question, we replace cross attention with naive MLP (neighbor-mlp prompt), which simply concatenates surrounding context embedding and utilizes MLP to aggregate information.
The results are illustrated in Fig~\ref{fig:bar_ablation_study}.
From the result, we have the following observation: (1) Replacing neighbor-aware prompt with point-wise prompt leads to performance decline, which reveals the effectiveness of the surrounding context. (2) We observe a sharp performance drop when using the neighbor-mlp prompt.
The result indicates that the surrounding context may introduce noise at the same time, and thus, an effective module is essential for capturing useful information within it.

\begin{table}[htbp]
\centering
\caption{Ablation study of key designs in geo-aware self-attention. 
}
\resizebox{1.0\linewidth}{!}{
\begin{tabular}{c|c|c|c|c|c|c}
\hline
\multirow{2}{*}{Method} & \multicolumn{3}{c|}{Recall} & \multicolumn{3}{c}{NDCG} \\ \cline{2-7}
& {@5} & {@10} & {@20} & {@5}  & {@10} & {@20} \\ \hline
Full Model & \textbf{0.3565}&\textbf{0.4563}&\textbf{0.5584}&\textbf{0.2032}&\textbf{0.2114}&\textbf{0.2151}\\ \hline
w/o Location Scores & 0.3476&0.4439&0.5229&0.1758&0.1847&0.1884 \\
w/o Location Gate & 0.3501&0.4489&0.5295&0.1810&0.1914&0.1950 \\ 
w/o Geo-aware Self-attention & 0.3315&0.4261&0.4989&0.1552&0.1640&0.1673\\ \hline
\end{tabular}
}
\label{table:ablation_1}
\end{table}

\textbf{Geo-aware self-attention.}
Two key designs of geo-aware self-attention are: (1) introducing location context similarity into attention scores (location scores) and (2) leverage user's real-time location as a gate (location gate).
To demonstrate the effectiveness of geo-aware self-attention, we carefully design three variants:
\begin{itemize}[leftmargin=*]
    \item \textit{w/o location scores}: We calculate attention scores without location context scores, i.e., removing the term $E^{lc}(E^{lc})^T$ in Eq~\ref{eq:location context scores}.
    \item \textit{w/o location gate}: We remove the location gate, i.e., Eq~\ref{eq:gate}.
    \item \textit{w/o geo-aware self-attention}: We replace geo-aware self-attention with vanilla self-attention, which can be considered as the combination of \textit{w/o location context scores} and \textit{w/o location gate}.
\end{itemize}
The results are summarized in Table~\ref{table:ablation_1}.
Compared to the full model, the performance decline of the three variants underscores the critical role of geo-aware self-attention and its key designs.

\begin{figure}[ht]
    \centering
    \includegraphics[width=0.95\linewidth]{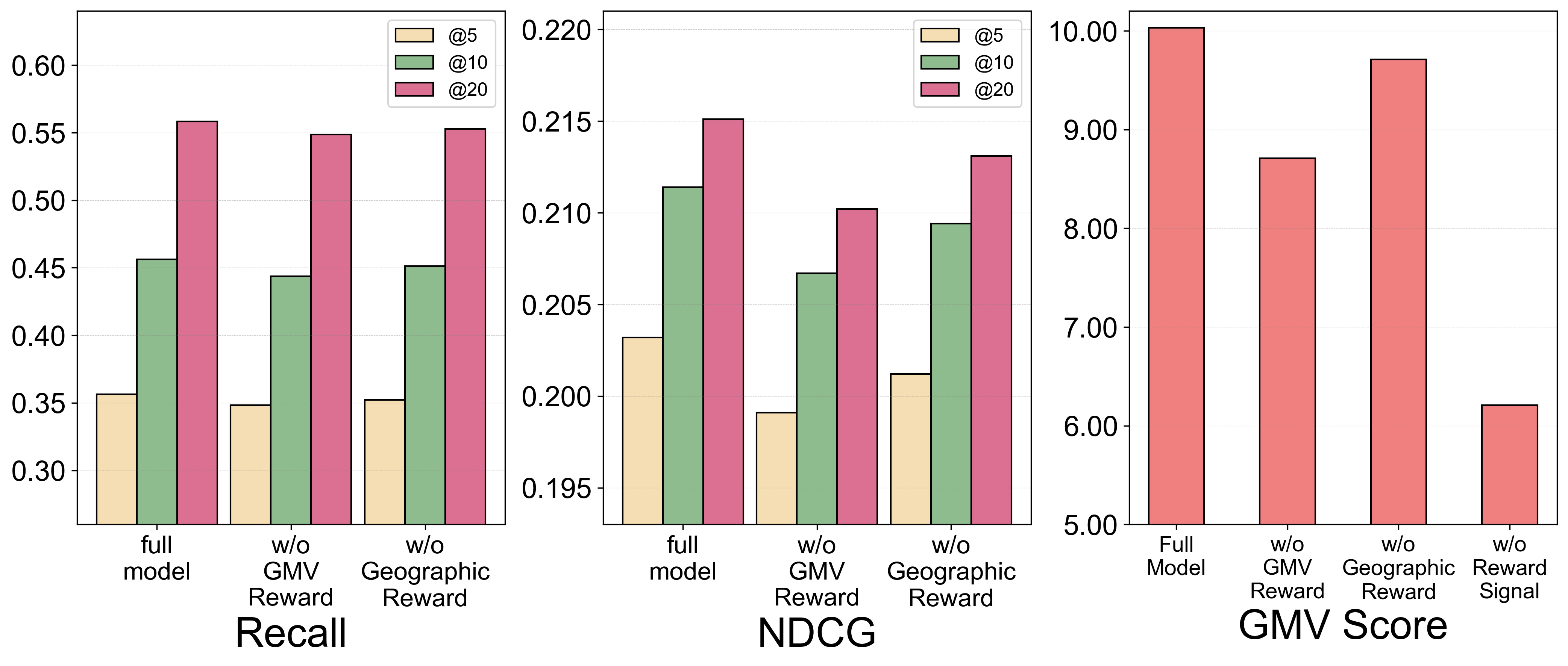}
    \caption{Conduct a comparative analysis of Recall, NDCG, and GMV metrics across varied reward signals.}
    \label{fig:rl}
\end{figure}

\textbf{Reward Signals.}
In the reinforcement learning, we use two rewards to align with multiple objectives.
We remove the geographic reward and the GMV reward separately to investigate their impact.
The results are showed in figure~\ref{fig:rl}.
Specifically, we remove geographic reward (\textit{w/o geographic reward}) and observe a decrease in recall and NDCG metrics.
This indicates that the geographic reward would force the model to consider the distance in spite of user preference, thus boosting the recommendation. 
In addition, we remove the GMV reward (\textit{w/o GMV reward}) and evaluate the performance using the three metrics (Recall, NDCG and GMV).
The result underscores that the GMV reward can not only boost the GMV objective but also benefit the recall objective.

\subsection{Hyperparameter Experiments (RQ3)}
\begin{figure}[ht]
    \centering
    \includegraphics[width=0.98\linewidth]{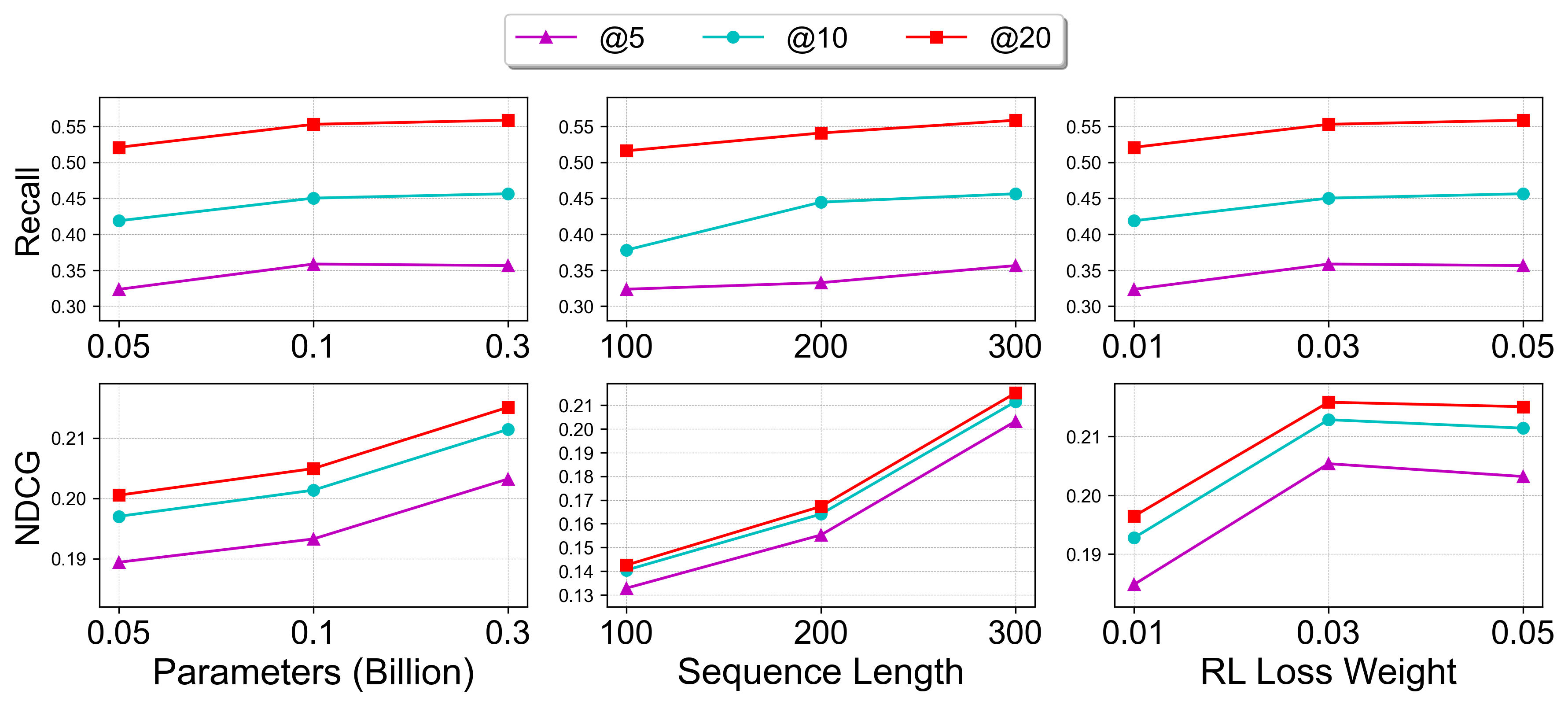}
    \caption{Hyperparameters include model parameters, sequence length and DPO loss weight $\lambda$. 
    }
    \label{fig:scaling}
\end{figure}

In this section, we conducted experiments to investigate the impact of each hyperparameters, including model size, sequence length and the DPO loss weight $\lambda$.
We change the hyperparameters one by one and the results are illustrated in Fig~\ref{fig:scaling}.
From the results, we have the following observation:
\begin{itemize}[leftmargin=*]
    \item In terms of model size and sequence length, we observe scaling laws. When scaling the model size from 0.05B (billion) to 0.1B, and then to 0.3B, we consistently observe an increase.
    When scaling the sequence length, we have the similar observation.
    \item From 0.05B to 0.3B, we achieve an average improvement of $6.96\%$ and $7.29\%$ in terms of recall and NDCG. When scaling sequence length from 100 to 300, we achieve an average improvement of $13.02\%$ and $51.24\%$ in terms of recall and NDCG.
    \item In terms of the loss weight, we find that the loss weight is sensitive, i.e., setting $\lambda=0.01$ is significantly worse than setting $\lambda=0.05$. Setting $\lambda=0.03$, both Recall@10 and Recall@20 exhibit significantly lower values compared to $\lambda=0.05$, whereas the NDCG metric demonstrates superior performance under $\lambda=0.03$. After a comprehensive evaluation of these trade-offs, we ultimately selected $\lambda=0.05$ as the final choice for our model.
\end{itemize}

\section{Online A/B Test}
To validate the effectiveness of our proposed model OneLoc, we conducted a one-week online A/B test on Kuaishou’s primary short-video recommendation scenario, which serves over 400 million daily active users. In this experiment, 10\% of the traffic was allocated to the treatment group using the end-to-end OneLoc system, while the control group retained the production-level multi-stage recommendation pipeline (including multi-channel retrieval, coarse/fine ranking, and link-specific refinements). Both groups shared identical candidate pools and system constraints to ensure fairness.

\begin{table}[t]
\centering
\caption{The absolute improvement of OneLoc compared
to the production-level  multi-stage system in the online A/B test.}
\label{tab:online_abtest}
\begin{tabular}{l|r}
\hline
{Online Metrics} & {OneLoc} \\
\hline
GMV & +21.016\% \\
Number of Orders & +17.891\% \\
Number of Paying Users in Local Services & +18.585\% \\
New Paying Users in Local Services & +23.027\% \\
\hline
\end{tabular}
\end{table}

As shown in Table \ref{tab:online_abtest}, OneLoc achieves significant improvements in key business objectives with $p < 0.01$ under two-tailed t-tests at $\alpha = 0.05$: GMV $+21.016\%$, order volume $+17.891\%$, and new local service buyers $+23.027\%$. These results demonstrate that OneLoc is able to surpass complex cascade systems in high-traffic industrial environments, especially in addressing cold-start and data sparsity challenges within local commerce.

\section{Related Work}

\subsection{POI Recommendation}
POI recommendation is similar with local life service recommendation (LLSR) since both need geographic information.
TPG~\cite{luo2023tpg} explicitly uses target times as prompts for a geography- sensitive recommendation.
STAN~\cite{luo2021stan} proposes a spatial-temporal attention mechanism to capture spatial-temporal relevance within POI trajectories.
Rotan~\cite{feng2024rotan} introduces a novel time-aware attention mechanism by representing time intervals as rotational position vectors. Thus, it is effective to capture time information within user behaviors.
LLM-Mob~\cite{wang2023LLM-Mob} introduce in-context learning to enhance next POI recommendation using historical and
contextual trajectories
NextLocLLM~\cite{liu2024nextlocllm} transforms ID prediction to coordinate prediction and injects spatial coordinates into LLM to enhance its understanding of spatial relationships between locations.
LLM4POI~\cite{li2024LLM4POI} transforms the next POI recommendation task into a question-answering task to effectively use abundant contextual information in POI recommendation.

In spite of the effectiveness of the above methods, they are all discriminative models, while generative recommendation has emerged as a new paradigm and showcased promising results. 

\subsection{Generative Recommendation}
Generative recommendation has garnered attention from both academia and industry.
TIGER~\cite{rajput2023tiger} is the first work to propose the generative recommendation framework using hierarchical semantic IDs encoded with RQ-VAE.
OneRec~\cite{deng2025onerec,zhou2025onerec-tech} further utilizes reinforcement learning for user preference learning and unifies the cascading framework with an end-to-end generative framework.
OneSug~\cite{guo2025onesug} proposes an end-to-end generative framework for e-commerce query suggestion.
COBRA~\cite{yang2025COBRA} proposes a coarse-to-fine framework that first generates semantic IDs and then generates dense vectors for retrieval.
LC-Rec~\cite{zheng2024lc-rec} aligns with collaborative filtering signals with multiple training tasks. 
LETTER~\cite{wang2024letter} improves the tokenizer by integrating hierarchical semantics, collaborative signals, and code assignment diversity.
ActionPiece~\cite{hou2025actionpiece} argues that the same action may have different meanings depending on its surrounding context and proposes a context-aware tokenizer.
EAGER~\cite{hong2025eager} aligns the linguistic semantics of pre-trained LLMs and the collaborative semantics in a non-intrusive manner.
Recently, GNPR-SID~\cite{wang2025poi_gr} migrates semantic-ID-baesd generative recommendation to the POI recommendation scenario.

However, prior generative recommendation work has not sufficiently explored the use of geographic information and the modeling of business goals. 
Thus, they are insufficient for a local life service system in the real industry scenario.

\section{Conclusion}
In this paper, we focus on our local life service scenario and propose a generative recommendation framework tailored for this scenario.
We propose three core modules to effectively model location information, including geo-aware semantic IDs, geo-aware self-attention, and neighbor-aware prompt.
To improve business objectives, we propose a reinforcement learning paradigm with a dual reward function including geographic reward and GMV reward.
Offline and online experiments have demonstrated the effectiveness of our method.
We also conducted rich experiments to investigate the impact of each component and hyperparameters.
Now OneLoc serves 400 million users daily in local life service of Kuaishou App and achieves $21.016\%$ and $17.891\%$ improvements in terms of GMV and orders numbers.
In the future, we will continue to explore some promising directions, including the scaling laws of model size and sequence length, and RL methods adapted to local life services.

\bibliographystyle{ACM-Reference-Format}
\bibliography{sample-base}




\end{document}